\documentclass[aps,prl,superscriptaddress,preprintnumbers,showpacs,twoside,twocolumn,floatfix]{revtex4}

\usepackage{bm}
\usepackage{graphicx}
\usepackage{amsmath}

\usepackage{amsmath, amssymb}
\usepackage{times}
\usepackage{mathtools}

\begin{document}

\title{Quantum Monte Carlo calculation of entanglement R\'enyi entropies for generic quantum systems}

\author{Stephan Humeniuk}
\affiliation{Laboratoire de Physique, CNRS UMR 5672, Ecole Normale Sup\'erieure de Lyon, Universit\'e de Lyon, 46 All\'ee d'Italie, 
Lyon, F-69364, France}
\affiliation{Institut de Ciencies Fotoniques i Optiques,  Av. C. F. Gauss, num. 3, 08860 Castelldefels, Spain}
\author{Tommaso Roscilde}
\affiliation{Laboratoire de Physique, CNRS UMR 5672, Ecole Normale Sup\'erieure de Lyon, Universit\'e de Lyon, 46 All\'ee d'Italie, 
Lyon, F-69364, France}

\begin{abstract}
We present a general scheme for the calculation of the R\'enyi entropy of a subsystem in quantum many-body models that can be efficiently simulated via quantum Monte Carlo. When the simulation is performed at very low temperature, the above approach delivers the entanglement 
R\'enyi entropy of the subsystem, and it allows to explore the crossover to the thermal R\'enyi entropy as the temperature is increased. We implement this scheme explicitly within the Stochastic Series expansion as well as within path-integral Monte Carlo, and apply it to quantum spin and quantum rotor models. In the case of quantum spins, we show that relevant models in two dimensions with reduced symmetry (XX model or hardcore bosons, transverse-field Ising model at the quantum critical point) exhibit an area law for the scaling of the entanglement entropy.    
\end{abstract}


\maketitle

Entanglement represents the unique correlation property of quantum states, without any classical counterpart, and as such it can play a fundamental role in our understanding of quantum many-body phases from the point of view of non-local correlations. 
 The most striking manifestation of entanglement in a quantum state $|\psi\rangle$ is represented by the mixed nature of the reduced density matrix $\rho_A$ describing a subsystem $A$ of a quantum many-body system, and defined as the partial trace of the total  density matrix $\rho = |\psi\rangle \langle \psi |$ on the complement $B$, $\rho_A = {\rm Tr}_B \rho$.  The mixedness of $\rho_A$ can be captured by any entropy estimator, the most common being the von-Neumann entropy $S^{\rm(vN)}_A = -{\rm Tr} \rho_A \log \rho_A$, but one can equivalently use its generalization, the R\'enyi entropy (RE) \cite{Renyi} 
 $S^{(\alpha)}_A = -\log[\rm{Tr}(\rho_A^{\alpha})]/(1-\alpha)$, which reduces to von-Neumann's in the limit $\alpha\to 1$. 
 The calculation of entanglement entropies in quantum many-body states appears as a formidable task, as it seems to imply the necessity to reconstruct the reduced density matrix of a subsystem $A$; this generally represents a hard problem unless $A$ contains very few degrees of freedom. In fact this task can be performed efficiently only in a few cases, including non-interacting bosons and fermions on a lattice \cite{Peschel03}. For the same models, considering fully connected $A$ regions
 (e.g. hypercubic ones) the scaling of the entanglement entropy with the linear size $l_A$ of the region can be calculated analytically in the asymptotic limit $l_A \to \infty$. Analytical and numerical calculations show that most models verify a so-called area law $S_A^{(\alpha)} \sim l_A^{D-1}$ in $D$ dimensions, except for critical models in $D=1$ and free fermions with a 
 ($D$-1)-dimensional Fermi surface, in which the area law is corrected by a multiplicative logarithmic term \cite{Eisertetal10}.    
Much less is known about models of interacting particles: indeed the reduced density matrix can be in principle reconstructed efficiently via density-matrix renormalization group (DMRG) only in one dimensional lattice systems, while for higher-dimensional systems the general method to reconstruct $\rho_A$ is via exact diagonalization, necessarily limited to small systems. It is also worth mentioning that the calculation of the R\'enyi entropy for $\alpha=2$ has been implemented for SU(2)-invariant lattice spin models via a projector Monte Carlo technique in Ref.~\onlinecite{Hastingsetal10}. 
Here we propose a new technique to calculate the R\'enyi entropy of a subsystem, valid for \emph{arbitrary} quantum many-body models which admit an efficient quantum Monte Carlo (QMC) solution of their equilibrium statistical properties. The basic idea is to perform the QMC simulation in an extended ensemble for $\alpha$ replicas of the system, treating the topology of the ($D$+1)-dimensional configurations generated by QMC as a dynamical variable. We demonstrate this approach for the calculation of the $\alpha=2$ R\'enyi entropy both at (physically) zero and finite temperature, for two-dimensional $S=1/2$ quantum spin models with low symmetry, as well as for the O(2) quantum rotor model in $D=1$.  
 
\begin{figure}[h]
\includegraphics[width=5cm]{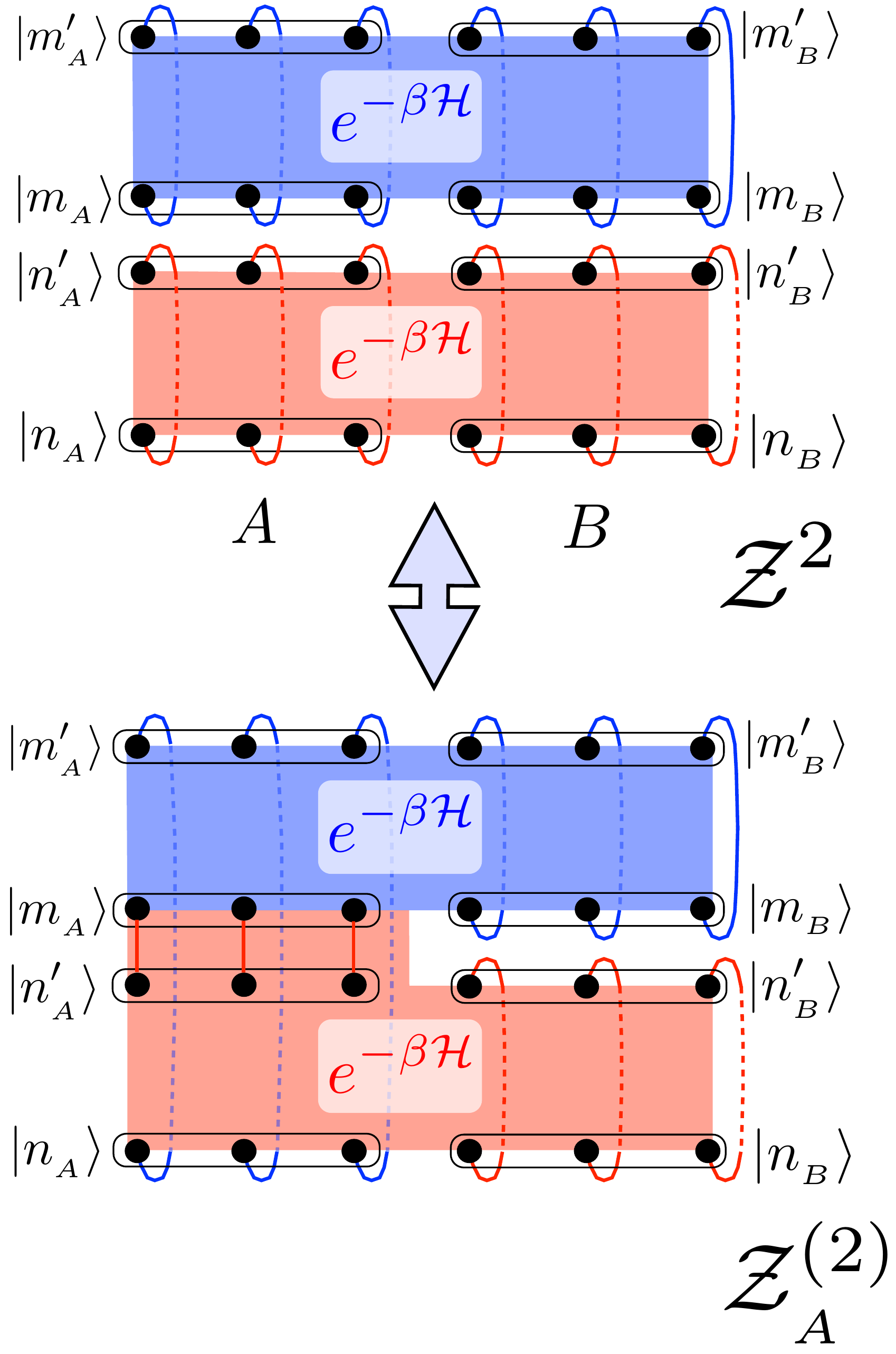}
\caption{Transition from the ${\cal Z}^2$ to the ${\cal Z}_A^{(2)}$ sector by redefinition of the topology of the simulation box in the additional dimension. The blue/red regions and lines are associated with the action of the imaginary-time propagator $e^{-\beta{\cal H}}$.} 
\label{f.sectors}
\end{figure}

Ref.~\onlinecite{CalabreseC04} has shown that, thanks to its trace structure, the R\'enyi entropy at finite temperature can be cast in the form of the logarithm of the ratio of partition functions $S^{\alpha} = \log{R_A^{(\alpha)}}/(1-\alpha)$ where $R_A^{(\alpha)} = {\cal Z}^{(\alpha)}_A/{\cal Z}^{\alpha}$; here 
${\cal Z}^{\alpha}  = \left[{\rm Tr}\left(e^{-\beta \cal H}\right)\right]^{\alpha}$  is the ordinary partition function for $\alpha$ replicas of the system, while ${\cal Z}^{(\alpha)}_A$ is a modified partition function for replicas which are ``glued" together in the region $A$. This is best seen in the simplest case $\alpha=2$, for which
\begin{equation}
{\cal Z}_A^{(2)} = 
\sum_{\mathclap{\substack{n_{A},m_{A} \\ n_{B},m_{B}}}}~~
\langle n_A n_B | e^{-\beta \cal H} | m_A n_B\rangle \langle m_A m_B | e^{-\beta \cal H} | n_A m_B\rangle~.
\label{e.ZA2}
\end{equation}
Here $|n_A n_B\rangle$ is an arbitrary basis of states which are factorized between the $A$- and $B$-region. 
Regarding $e^{-\beta \cal H}$ as the imaginary-time propagator, Eq.~\eqref{e.ZA2} describes a cyclic propagation for a time $2\beta$ of the $A$-region  state, and two independent cyclic propagations for a time $\beta$ of the $B$-region state - as sketched in Fig.~\ref{f.sectors}. On the other hand ${\cal Z}^2$ describes two independent propagations for a time $\beta$.
   
The statistical mechanics formulation of the R\'enyi entropy has been remarkably exploited in Ref.~\onlinecite{CalabreseC04} for the calculation of the entanglement entropies for conformal field theories (CFT); more recently Refs.~\onlinecite{Melkoetal10, Isakovetal11,Singhetal11} have implemented a quantum Monte Carlo calculation of both ${\cal Z}$ and  ${\cal Z}_A^{(\alpha)}$ separately via direct thermodynamic integration of the energy curve $E(\beta')$ over the interval $[0,\beta]$ to obtain the finite-temperature R\'enyi entropy $S_A^{(\alpha)}(\beta)$. This technique, while being very general, appears to be technically limited to finite temperatures, given that the statistical error accumulated in the thermodynamic integration from infinite temperature down to the temperature of interest grows significantly at low $T$ \cite{Melkoetal10, Isakovetal11}. 
Here we propose an alternative QMC approach which cures the above limitation, allowing to systematically calculate the R\'enyi entropy of a subsystem with a \emph{single} simulation at the temperature of interest, performed within an extended ensemble. 
The central idea of our approach is that the ratio of two partition functions can be generally estimated with Monte Carlo by performing a simulation in an ensemble 
which is the union of the two, ${\cal Z}^2 \cup {\cal Z}_A^{(2)}$. Whichever quantum Monte Carlo approach is used for the estimation of the equilibrium statistical properties of the Hamiltonian ${\cal H}$, it should allow to write ${\cal Z}_A^{(2)}$ in the form
\begin{equation}
{\cal Z}_A^{(2)} = \sum_{\cal C} w_A\left({\cal C}\right)
\end{equation}
where ${\cal C} = \left(n_A,n_B;m_A,m_B;{\cal P}\right)$ is a QMC configuration, in which 
the state  $|n_A\rangle$ is propagated to 
$|m_A\rangle$ and then back to itself in the region $A$, while the states $|n_B\rangle$ and $|m_B\rangle$ are 
propagated onto themselves independently, and the propagation scheme is represented by ${\cal P}$: ${\cal P}$ is generally a 
path in the computational basis $|\psi(\tau)\rangle = |\psi_A(\tau)\rangle |\psi_B(\tau)\rangle$ parametrized by the (continuous) imaginary time $\tau \in [0,\beta]$
as in path-integral Monte Carlo (PIMC) \cite{Ceperley95}, or by the propagation step index $\tau=p$, associated with a string of bond operators, as in 
Stochastic Series Expansion (SSE) \cite{SyljuasenS02}. Within the above notation, ${\cal Z}^2 = {\cal Z}_{A=\varnothing}^{(2)}$.
$w_A$ is the statistical weight of a configuration;
the Hamiltonian ${\cal H}$ lends itself to an efficient QMC simulation if $w_A\geq 0$ for all configurations, and if the weights
$w_{A}$ can be calculated efficiently.

Our method is then based on constructing a simulation which moves dynamically between the ${\cal Z}^2$ ensemble and the ${\cal Z}^{(2)}_A$
while respecting detailed balance condition. The move from one ensemble to another can be performed with Metropolis probability
\begin{equation}
P\left({\cal Z}^2 \to {\cal Z}^{(2)}_A\right) = \min\left(1,\frac{w_A({\cal C})}{w_{A={\varnothing}}({\cal C})}\right)
\label{e.rate}
\end{equation}
and viceversa for the reverse move. The partition function ratio $R_A^{(2)}$ is then simply estimated as 
\begin{equation}
R_A^{(2)} = \left \langle \frac{N_A}{N_{A=\varnothing}} \right\rangle_{\rm MC}
\end{equation}
where $N_A$ is the number of MC steps in the ensemble with a given region $A$, and $\langle...\rangle_{\rm MC}$ is the Monte Carlo average.
A straightforward generalization of the above formulas is possible for $\alpha>2$. We would like to point out that an analogous extended-ensemble 
QMC scheme is the one defining the QMC estimator for observables which are off-diagonal in the computational basis \cite{Boninsegnietal06}. 

In practice, for a simulation on discrete degrees of freedom on a lattice - \emph{e.g.} quantum spins, lattice gases - the weights $w_A({\cal C})$
and $w_{A={\varnothing}}({\cal C})$ cannot be simultaneously non-vanishing unless the condition $|n_A\rangle = |m_A\rangle$ is satisfied, in which
case the transition in the propagation topology (Fig.~\ref{f.sectors}) is microcanonical, namely $w_A({\cal C}) = w_{A=\varnothing}({\cal C})$.
In the case of continuous lattice variables - \emph{e.g.} quantum rotors - or of particles in continuous space, the ``rewiring" of worldlines demanded by 
the transition between the ensembles can be in principle always performed, although it will have an acceptance rate which is low if the configurations
$|n_A\rangle$ and $|m_A\rangle$ are very different. Assuming to use a PIMC scheme in which the imaginary time is discretized in steps
$\Delta\tau$, and indicating with ${\cal H}_{\rm OD}$ the part of the Hamiltonian which is off-diagonal in the computational basis, one has that
the ratio $w_A({\cal C}) /w_{A={\varnothing}}({\cal C})$ takes the expression 
\begin{equation} 
\frac{\langle n_A'n_B'|e^{-\Delta\tau{\cal H}_{\rm OD}}|m_A n_B\rangle \langle m_A'm_B'|e^{-\Delta\tau{\cal H}_{\rm OD}}|n_A m_B\rangle}
{\langle n_A'n_B'|e^{-\Delta\tau{\cal H}_{\rm OD}}|n_A n_B\rangle \langle m_A'm_B'|e^{-\Delta\tau{\cal H}_{\rm OD}}|m_A m_B\rangle}~.
\end{equation}
As indicated in Fig.~\ref{f.sectors}, the primed configurations are those which are connected to the un-primed ones by a single propagation step.

\begin{figure}[h]
\includegraphics[width=6cm]{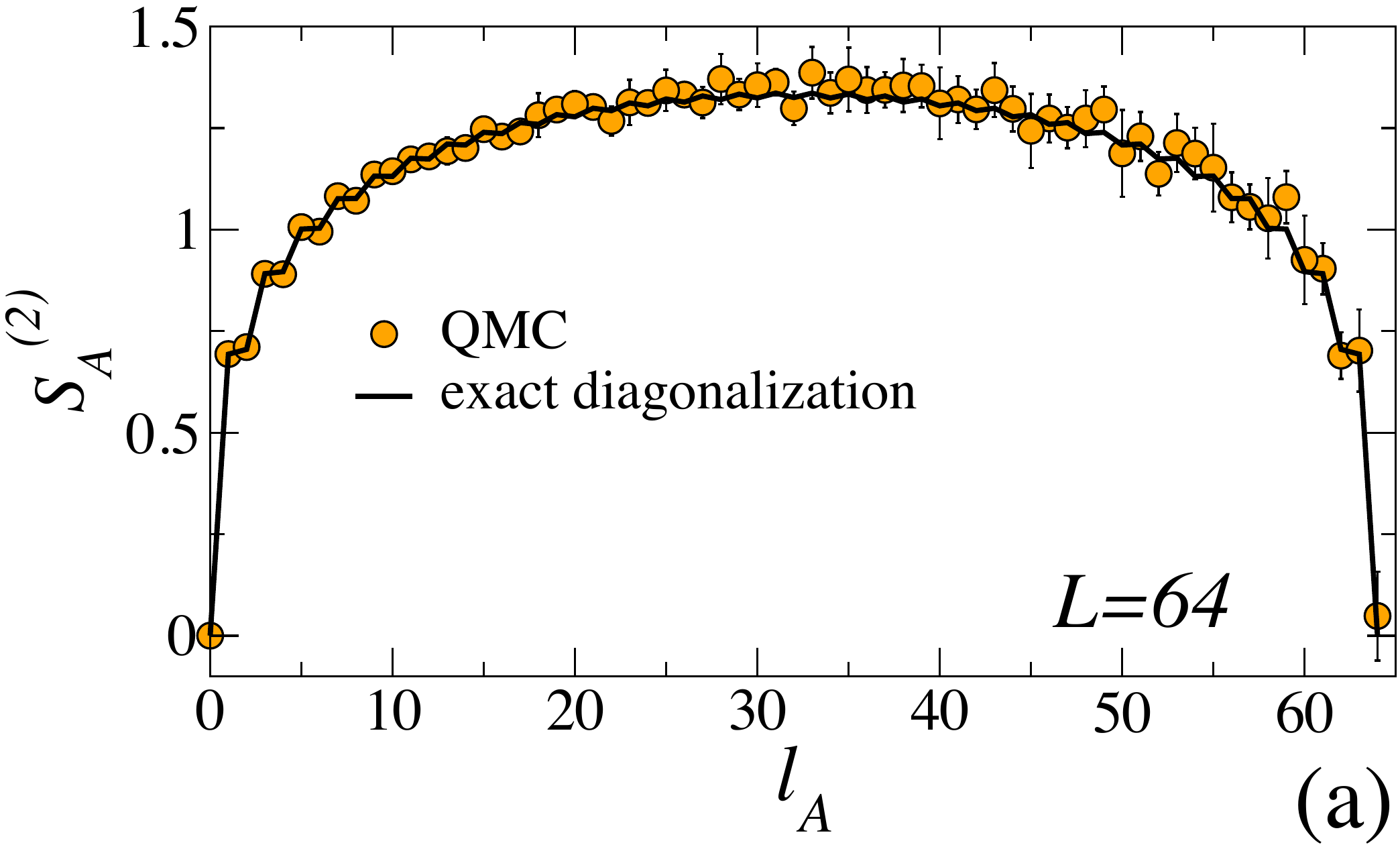} \\
\includegraphics[width=6cm]{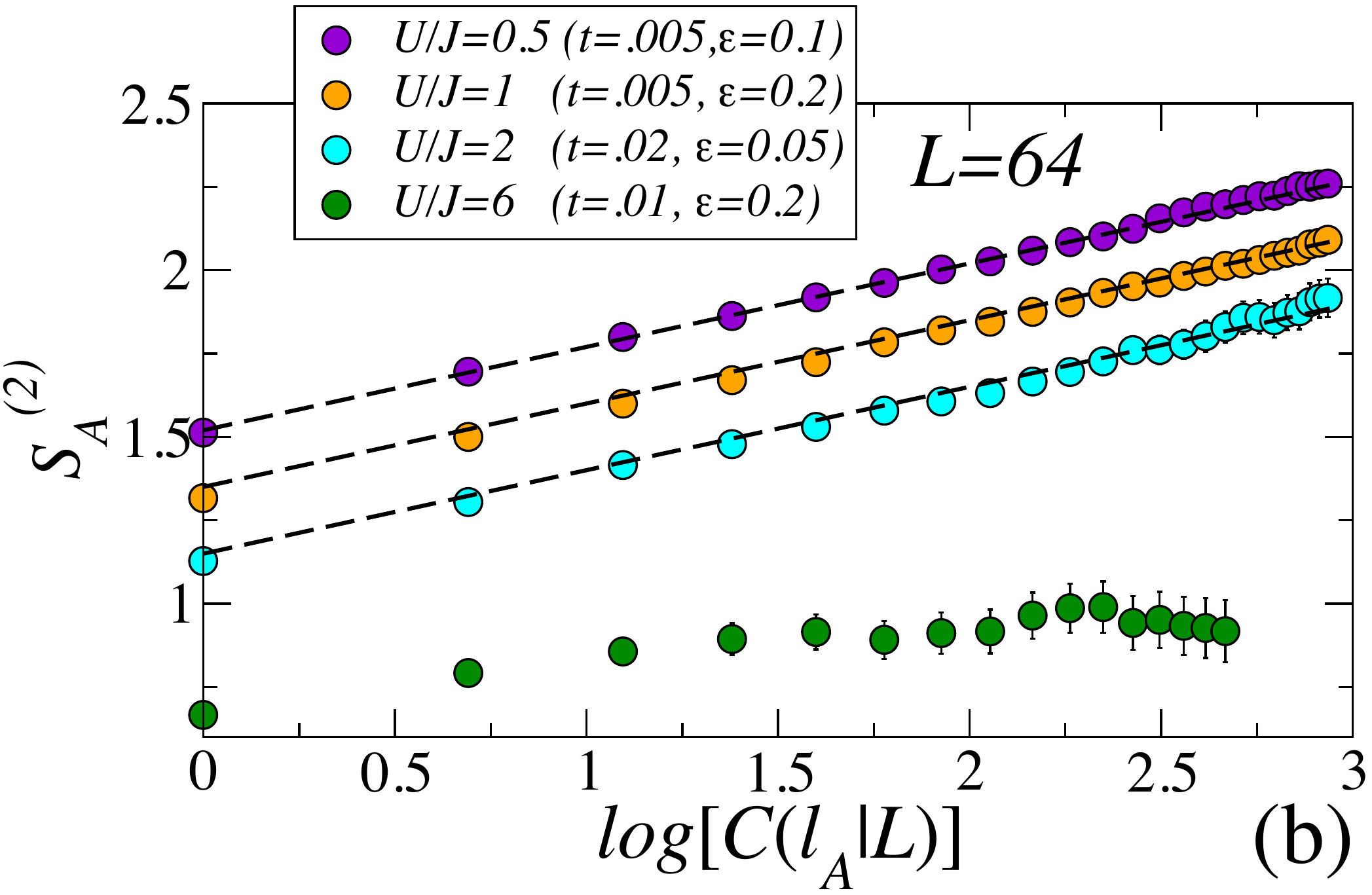}
\caption{Upper panel: Entanglement entropy of the 1$D$ XX chain; the solid line corresponds to exact diagonalization. Lower panel: Entanglement entropy of the 1$D$ O(2) quantum rotor model. Here $L=64$, $t = k_B T/ J$, and $\epsilon = \Delta\tau U$; increments $\Delta l = 1$ are used; the dashed lines correspond to fits to the CFT prediction $(1/4)\log[C(l_A|L)] + s_1$.} 
\label{f.1DXX}
\end{figure}

 The above scheme provides an efficient estimate of $R_A^{(\alpha)}$, and therefore of $S_A^{(\alpha)}$, by performing a single simulation at the temperature of interest. When the temperature is chosen to be so low as to remove thermal effects on a finite-size simulation box, one can gain access to the entanglement R\'enyi entropy.   
 Unless otherwise specified, in the following we will show simulation results for the general case of an XYZ Hamiltonian in a field
 \begin{equation}
 {\cal H} = J \sum_{\langle ij \rangle} \left(S_i^x S_j^x + \Delta_y S_i^y S_j^y + \Delta_z S_i^z S_j^z \right) - H \sum_i S_i^z~
 \end{equation}
 where $S_i^{\alpha}$ are $S=1/2$ spin operators, $J>0$, and $\langle ij \rangle$ indicates a pair of nearest neighbors on a $D$-dimensional hypercubic lattice.  
A validation of our approach comes from the comparison with exact results in $D=1$, which are available \emph{e.g.} for the case of the XX model 
($\Delta_y = 1$, $\Delta_z = 0$); such a model admits a mapping onto a system of free fermions \cite{LiebSM61}, whose entanglement properties can be calculated  from the knowledge of two-point correlations \cite{Peschel03}. Fig.~\ref{f.1DXX}(a) shows the data for a $L=64$ chain with periodic boundary conditions, simulated with the SSE algorithm; very good agreement is found between the exact results and QMC results at an inverse temperature $\beta J = 200$. 
In principle the whole $S_A^{(2)}$ curve as a function of $l_A$ can be obtained by performing simulations in the joint ${\cal Z}^2 \cup {\cal Z}_A^{(2)}$ ensemble; in practice, nonetheless, the transition rate between the two ensembles is strongly suppressed when the size of $A$ grows, given that the condition 
$|n_A\rangle =|m_A\rangle$ is increasingly hard to satisfy. This aspect reflects the fact that the ratio $R_A^{(2)}$ estimated in the simulation decreases exponentially with $S_A^{(2)}$,
 $R_A^{(2)} = \exp(-S_A^{(2)})$. In the $D=1$ case in question, in which $S_A^{(\alpha)} \approx (c/6)(1+1/\alpha) \log l_A$ \cite{CalabreseC04},  
$R_A^{(2)} \approx l_A^{-c/4}$; in the case in which an area law holds, one has an even more serious decrease $R_A^{(2)} \approx \exp(-\tilde{b}~l_A^{D-1})$.
Hence the events that the simulation has to count become increasingly rare, so that the simulation length should na\"ively scale as $(R_A^{(2)})^{-1}$.
To cure this problem we use the \emph{increment trick} from Ref.~\onlinecite{Hastingsetal10}, by formally rewriting $R_A^{(2)}$ as
\begin{equation}
R_A^{(2)} = \prod_{i=0}^{N-1} R^{(2)}_{A_i,A_{i+1}}~~~~~~ R^{(2)}_{A_i,A_{i+1}} = \frac{{\cal Z}_{A_{i+1}}^{(2)}}{{\cal Z}_{A_{i}}^{(2)}}
\end{equation}
 where $A_i$ is a sequence of $N$ blocks of increasing size such that $A_0 = \varnothing$ and $A_N=A$. Each of the ratios $R^{(2)}_{A_i,A_{i+1}}$ can be 
 estimated efficiently, as it represents the ratio between the partition functions of systems which are $2\beta$-periodic on regions $A_i$ and $A_{i+1}$ chosen so as to differ only by a few sites (or by a few interparticle spacings in continuum space). 
The R\'enyi entropy is then the sum of contributions from the successive increments $\Delta A_i$ that lead from $\varnothing$ to $A$, 
$S_A^{(2)} = \sum S_{A_i,A_{i+1}}^{(2)}$ where $S_{A_i,A_{i+1}}^{(2)} = - \log R^{(2)}_{A_i,A_{i+1}}$. While too large increments  $\Delta A_i$ give rise to inefficient estimates of  the corresponding ratios $R^{(2)}_{A_i,A_{i+1}}$, too small ones lead to a sizable accumulated error on the sum; yet an optimal size of the increment can be found minimizing the final error on $S_A^{(2)}$. In Fig.~\ref{f.1DXX}(a) we have used linear increments of size $\Delta A_i = \Delta l = 5$. Nonetheless it can still be seen that the precision of the results is not optimal for $l_A \gtrsim 20$. This is a result of the slow increase of entanglement entropy in 1$D$ systems (especially for $l_A \approx L/2$): if the ratios $R^{(2)}$ are known with a given relative error $\epsilon_R = \Delta R^{(2)}/R^{(2)}$, the corresponding entanglement increment $S^{(2)}$ has a relative error $\epsilon_S = \epsilon_R/S^{(2)}$, which can be much bigger than $\epsilon_R$ when the increment is small. 
This means that the QMC technique enjoys a faster scaling of the entropy, as found \emph{e.g.} in 2$D$ systems or at finite temperature; as we will see, the quality of the 2$D$ data is significantly better. 
\begin{figure}[h]
\includegraphics[width=7cm]{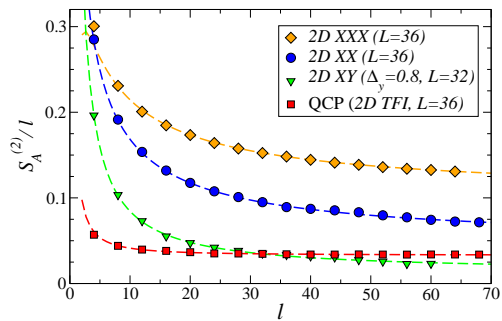}
\null\vspace*{-.8cm}
\caption{Entanglement entropy 2$D$ spin models with various symmetries; the error bars are smaller than the symbol sizes.The dashed lines represent fits to the equation in the main text. } 
\label{f.2D}
\end{figure}

 Having validated the approach against exact results, we can apply it to yet unexplored models. To demonstrate the versatility of the QMC RE estimator, we apply it to the study of a model with continuous quantum variables, namely the 1$D$ O(2) quantum rotor model ${\cal H} = -2J \sum_{\langle ij \rangle} \cos(\phi_i - \phi_j) - \frac{U}{2}\sum_i \partial^2/\partial\phi^2_i$, in which $\phi_i \in [0,2\pi]$. This model represents an approximation to the Bose-Hubbard model with hopping $J$ and repulsion $U$ for large integer filling, and it exhibits a superfluid-insulator quantum phase transition for increasing $U/J$. 
Such a model can be studied via PIMC \cite{Wallinetal94} with discretized imaginary time (in steps $\Delta\tau$). Fig.~\ref{f.1DXX}(b) shows the RE for a chain of $L=64$ sites at variable $U/J$; we observe that for sufficiently small $U/J$ the RE obeys the CFT prediction $S_A^{(2)} = (\bar{c}/4) \log[C(l_A|L)] + s_1$ where $\bar{c}=1$, $C(x|L) = L/\pi\sin(\pi x/L)$, and $s_1=s_1(U/J)$ is a constant dependent on the Hamiltonian parameters. On the other hand for large $U/J$ the CFT prediction is no longer verified, as the system enters an insulating gapped phase. 
  
 We then move to 2$D$ systems, and consider three representatives of the three symmetry sectors of the XYZ model in zero field, namely the case of a SU(2) invariant Heisenberg (or XXX) model ($\Delta_y = \Delta_z = 1$), the case of a U(1) symmetric XX model ($\Delta_y = 1$, $\Delta_z = 0$), and the case of the Z$_2$ symmetric anisotropic XY model ($\Delta_y = 0.8$, $\Delta_z=0$). In all three cases we consider $A$ regions with a square geometry $l_A\times l_A$, grown in linear increments of (typically) 5 sites, and we plot the data as a function of the region boundary $l = 4(l_A-1)$. The simulations have been performed with the SSE algorithm on lattices with $L\times L$ size up to $L=36$, and at a temperature $\beta J \approx L$ guaranteeing the removal of thermal contributions.  
\begin{figure}[h]
\includegraphics[width=7cm]{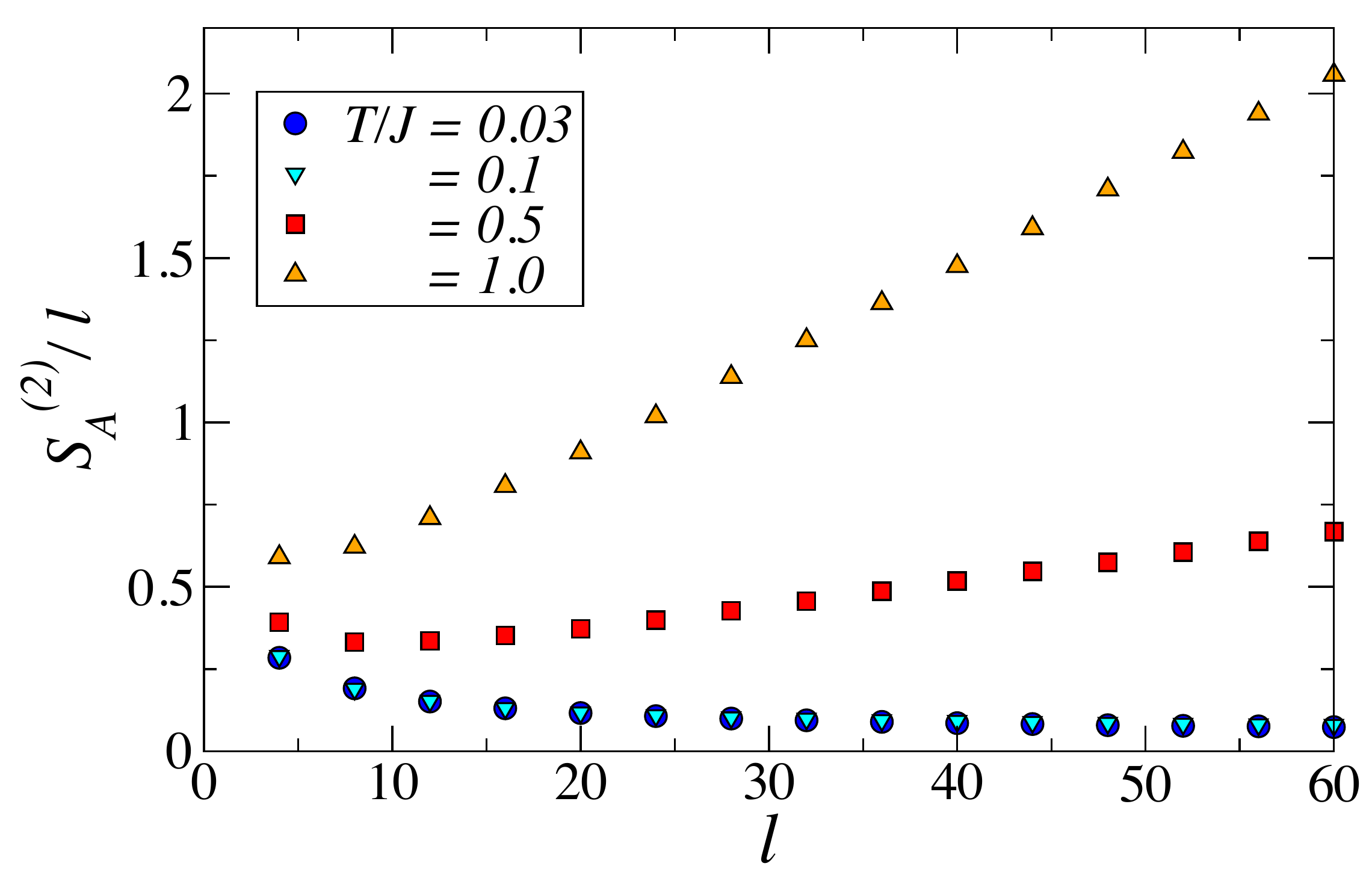}
\caption{R\'enyi entropy for the 2$D$ XX model at increasing temperature. Error bars are smaller than the symbol size.} 
\label{f.2DXX}
\end{figure}
The case of the XXX model has been previously investigated in Refs.~\cite{Hastingsetal10, Kallinetal11} via projector QMC, and we confirm their finding of an area law scaling of entanglement entropy. As clearly shown in Fig.~\ref{f.2D}, an area law is also observed for the other two models with reduced symmetry: in all three cases the scaling of the R\'enyi entropy is very well fitted by an area law plus subleading corrections, $f(l) = b ~l + c \log l + d$;
the fit coefficients (obtained by discarding data with $l< l_{\rm min}$) are reported in Table \ref{t.2D} \cite{supplemental}.   
In particular, the coefficient $b$ of the dominant area-law term decreases systematically as the symmetry of the model is decreased; this is consistent with the picture that a lower symmetry confines quantum fluctuations to a restricted region of spin space, thereby lowering entanglement properties. 
\begin{table}
\begin{tabular}{|l|c|c|c|r|}
  \hline
 model & $b$ & $c$ & $d$ & $l_{\rm min}$\\
  \hline
  2$D$ XXX ($\Delta_y = \Delta_z = 1$) & 0.099(1) & 0.48(4) &  0.05(10)  & 16 \\
  2$D$ XX ($\Delta_y = 1$, $\Delta_z = 0$) & 0.045(1) & 0.31(3) & 0.52(5)  & 8 \\
  2$D$ XY ($\Delta_y = 0.8$, $\Delta_z = 0$)&  0.013(1) &  -0.02(2) & 0.76(2) & 8 \\
  2$D$ TFI ($\Delta_y = \Delta_z = 0$) - QCP & 0.0332(4) & -0.03(1) & 0.15(2) &  8\\
  \hline
\end{tabular}
\caption{Fit coefficients for the three models investigated in Fig.~\ref{f.2D}.}
\label{t.2D}
\end{table}

The 2$D$ XX model maps onto hardcore bosons, and it is directly relevant \emph{e.g.} to current cold-atom experiments. To make contact with a more realistic experimental situation, we have studied the effect of an increasing temperature on the scaling of the R\'enyi entropy, an aspect which can be quite naturally investigated with finite-$T$ QMC.  
Given that thermal entropies are extensive, on general grounds one expects finite temperatures to introduce a \emph{volume} law in the scaling, namely a $a(T) l_A^D$ term fatally masking the area-law term. Nonetheless the growth of the volume-law coefficient with temperature, $a(T)$, appears fairly slow: as shown in Fig.~\ref{f.2DXX} an area law is also observed at a moderate, finite temperature, $T/J = 0.1$, for the block sizes considered here (the biggest being $l_A = 16$), implying that $a(T) l_A^D \ll 4 b (l_A-1) $. On the other hand, at a temperature $T/J = 0.5$ and higher the volume law term dominates already for small block sizes. These results point at the fact that area laws of R\'enyi entropy are observable even at finite temperature and for moderate block sizes, which are indeed relevant for experiments. 

 We conclude our discussion of $D=2$ models with the case of the 2$D$ transverse-field Ising (TFI) model, $\Delta_y = \Delta_z=0$, which displays a quantum critical point (QCP) at $H_c \approx 1.52 J$ \cite{BloteD02}. Our approach enables us to investigate this two-dimensional quantum critical system  in search for special entanglement signatures. As shown in Fig.~\ref{f.2D} and in Table \ref{t.2D}, we observe that the entanglement RE obeys an area law with a negative logarithmic correction and a positive additive constant. These findings are in quantitative agreement with recent field-theory results for a QCP with dynamical critical exponent $z=1$, predicting universal negative logarithmic corrections coming from corners of the $A$ region \cite{CasiniH07, supplemental}, and a universal ($\alpha$-dependent) additive constant in $D=2$ \cite{Metlitskietal09}.  This shows that QMC simulations can quantitatively extract the subleading corrections; their universality can be directly tested by investigating different microscopic models exhibiting QCPs in the same universality class.  
 
 In conclusion we have demonstrated a simple approach to incorporate an estimator of subsystem Renyi entropies into any finite-temperature quantum Monte Carlo scheme. This approach complements the estimator developed within the projector-QMC scheme \cite{Hastingsetal10}, and it paves the way for a systematic investigation of entanglement entropies in a large variety of interacting quantum systems in arbitrary dimensions, such as quantum fluids, quantum spin systems, $O(N)$ quantum rotor models, quantum field theories etc., as long as they are accessible to a QMC study.
 Contrary to DMRG or to variational methods based on tensor-network states, QMC simulations are completely unbiased with respect to the scaling of entanglement, and the QMC entanglement estimator actually performs better the faster the entanglement grows with the subsystem size.     
 Moreover QMC represents a natural platform to investigate the statistics of local quantum fluctuations in realistic systems, in the attempt to relate measurable fluctuation properties with entanglement properties \cite{Songetal11}. 

T. R. acknowledges fruitful discussions with J. I. Cirac and R. Melko which have sparked the present study.

\section{Supplementary Material}

\begin{figure}[h]
\includegraphics[width=9cm]{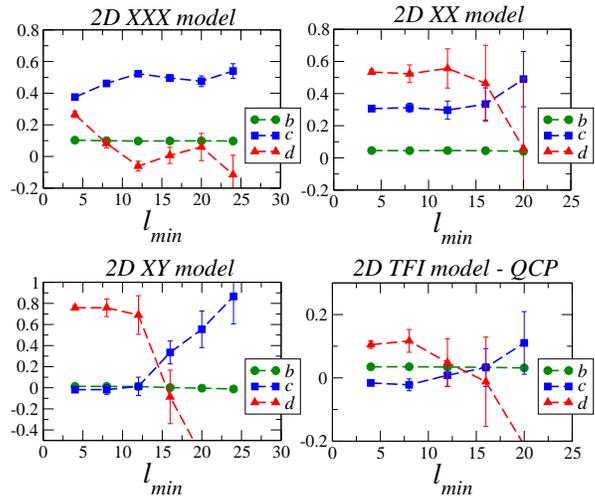}
\caption{Evolution of the fit coefficients of $S_A^{(2)}$ when increasing the lower bound of the fit region $[l_{\rm min},4(L/2-1)]$.} 
\label{f.2Dfits}
\end{figure}

 Here we provide a more detailed discussion of the coefficients of the R\'enyi entropy scaling of a square subsystem embedded in a 2$D$ quantum spin system. 
 We fit our QMC data for 2$D$ quantum spin models to the form $f(l) = b~l + c \log l + d$. On a $L\times L$ lattice, the fits are performed over a region of boundary sizes $[l_{\rm min},4(L/2-1)]$ whose lower bound $l_{\rm min}$ is gradually grown to check convergence. Fig.~\ref{f.2Dfits} shows the fit coefficients as a function of $l_{\rm min}$. As a general criterion, if convergence of the fitting parameters within the error bar is achieved for a given  $l^*_{\rm min}$, we choose as best fit parameters (shown in Table \ref{t.2D}) the ones corresponding to $l^*_{\rm min} - 4$ (given that they are consistent with the  $l^*_{\rm min}$ values and have smaller error bars). In general we observe that the coefficient of the area law $b$ is very stable to variations of $l_{\rm min}$, while shrinking too much the fitting region leads to a transition in the coefficients of the subleading terms $c$ and $d$. Nonetheless we observe that convergence in the fitting coefficients is achieved before the transition; we observe that the transitions are systematically accompanied by a degradation in the precision of the resulting fit coefficients, and we argue that they can be attributed to the limited data sets that are left to fit if $l_{\rm min}$ grows too big - such limited data sets can only provide reliably the coefficient of the leading term, but hardly those of the subleading ones.  
 
  Recent field-theoretical studies \cite{sCasiniH07, sCasiniH09, sMetlitskietal09, sMetlitskiG11} have pointed out that several two-dimensional quantum systems should exhibit a dominant area-law scaling of the entanglement entropy with a non-universal coefficient $b$ - given that such coefficient would depend on the short-distance cutoff related to the details of the microscopic model of origin.  On the other hand, the subleading logarithmic and constant terms can indeed be cutoff independent and universal. In the case of $z=1$ quantum critical points (QCPs) \cite{sMetlitskietal09}, if the $A$ region has a smooth boundary one expects that $b=0$ and an additive universal constant $d$, dependent on the index $\alpha$ of the considered Renyi entropy. Our finding for the QCP of the 2$D$ transverse field Ising (TFI) model is that $d$ is indeed finite and positive for $\alpha=2$; in the case $\alpha=1$ (von-Neumann's entropy) Ref.~\onlinecite{sTagliacozzoetal09} estimates as well a positive additive constant. On the other hand, we also find a finite logarithmic correction with a negative coefficient, 
  $c = -0.03(1)$. This is not at all surprising, given that the long wavelength properties of the QCP in the 2$D$ TFI model should be represented by a free relativistic theory. For such a theory Ref.~\onlinecite{sCasiniH07} predicts that a region $A$ with corners on the boundary will acquire universal negative logarithmic contributions; in particular each corner should contribute a term $\approx - 0.0062$ to the $c$ coefficient, which for the 4 corners of square $A$ regions provides a value in quantitative agreement with our estimate of the $c$ coefficient. 
 
 If the ground state has a finite correlation length (as it is the case for the 2$D$ anisotropic XY model), one expects a correlation-length dependent additive constant $d$ \cite{sMetlitskietal09}; the existing predictions of logarithmic corrections do not apply to this case. Indeed we find a sizable additive constant, and a logarithmic correction which is consistent with zero.
  Finally, in the case of the 2$D$ XX and XXX model, the ground state has an infinite correlation length and it develops long-range order in the thermodynamic limit. 
 For the case of the 2$D$ XXX model, fits to projector QMC data have been performed in Ref.~\onlinecite{sKallinetal11} using the function $f'(l_A) = 4a' l_A + c' \log(4l_A) + d'$ of the boundary size estimated as $4 l_A$ (this estimate double-counts the corner spins). When fitted to the $f'$ function our data deliver coefficients which are indeed in agreement with the ones quoted in Ref.~\onlinecite{sKallinetal11}.
  
  Both the XX and the XXX model have linearly dispersing gapless Goldstone modes (two for the XXX model, and one for the XX model), each described in the long-wavelength limit by a free relativistic theory. Following Ref.~\onlinecite{sCasiniH07} one would expect negative logarithmic corrections coming from corners, but in fact our results point at a positive $c$ coefficient for both models. This result is consistent with what was initially found numerically in Ref.~\onlinecite{sKallinetal11}, where positive logarithmic corrections have been shown to exist for the XXX model even in absence of corners.  
Prompted by the results of  Ref.~\onlinecite{sKallinetal11},  Ref.~\onlinecite{sMetlitskiG11} has recently predicted that in systems exhibiting spontaneous symmetry breaking in the thermodynamic limit, one should expect a positive logarithmic correction with a coefficient $c$ which takes the simple form $N_G(D-1)/2$ where $N_G$ is the number of Goldstone modes. This would imply that $c = 1$ for the 2$D$ XXX model and $c=1/2$ for the 2$D$ XX model. Our observation is not consistent with this prediction, even when taking into account negative logarithmic contributions coming from corners. Nonetheless we observe that the coefficient of the XXX model is significantly larger than that for the XX model. 
 One might argue that a possible source of discrepancy between our results and those of Ref.~\cite{sMetlitskiG11} stems from the finite-temperature nature of our data. Indeed a temperature $T \sim L^{-1}$ (at which our simulations are conducted) is sufficiently low to eliminate Goldstone-mode excitations, but not to eliminate the thermal occupation of the low-lying tower-of-states excitations \cite{Anderson1983} (to eliminate those states one would need a prohibitively low temperature, $T \sim L^{-2}$); yet Ref.~\onlinecite{sMetlitskiG11} suggests that in the case $L^{-2} \ll T \ll L^{-1}$ their prediction should still hold. A further source of discrepancy could be the fact that our finite-size results fail to correctly capture the behavior of the subleading terms in the limit $l\to\infty$. Future larger-scale simulations should be able to clarify this issue.

\end{document}